\documentstyle[prl,aps,multicol]{revtex}

\begin{document}
\draft

\def\beq{\begin{equation}}
\def\eeq{\end{equation}}
\def\eeql#1{\label{#1} \end{equation}}
\def\bea{\begin{eqnarray}}
\def\eea{\end{eqnarray}}
\def\eeal#1{\label{#1} \end{eqnarray}}
\def\phs{{\vphantom{*}}}
\def\hn{\mskip-0.5\thinmuskip}
\def\hp{\mskip0.5\thinmuskip}
\renewcommand{\vec}{\bbox}
\def\im{\mathop{\rm Im}\nolimits}
\def\re{\mathop{\rm Re}\nolimits}
\def\ds{\displaystyle}
\def\ts{\textstyle}
\def\half{{\ts\frac{1}{2}}}
\def\abs#1{\mathopen|#1\mathclose|}
\def\mL{\mu_{\rm l}}
\def\mR{\mu_{\rm r}}
\def\aL{\alpha_{\rm l}}
\def\aR{\alpha_{\rm r}}

\title{Quantum-efficient charge detection using a single-electron transistor}

\author{Alec Maassen van den Brink\thanks{Electronic address: {\tt alec@felix.physics.sunysb.edu}}}

\address{Department of Physics and Astronomy, SUNY at Stony Brook, Stony Brook, NY 11794-3800, USA}

\date{\today}
\maketitle

\begin{abstract}

We evaluate the detector nonideality (and energy sensitivity) of a normal-state single-electron transistor (SET) in the cotunneling regime in a two-charge-state approximation. For small conductances and at zero temperature, the SET's performance as a charge-qubit readout device is characterized by a universal one-parameter function. The result shows that near-ideal, quantum-limited measurement is possible for a wide range of small bias voltages. However, near the threshold voltage for crossover to sequential tunneling, the device becomes strongly nonideal. The (symmetrized) current-charge cross-correlation vanishes for low frequencies, causing two different definitions of detector nonideality to agree. Interpretations of these findings are discussed.

\end{abstract}

\pacs{PACS numbers: 73.23.Hk
, 03.67.-a
, 05.40.-a
, 07.50.Ls
}
 
\begin{multicols}{2}


\bigskip\noindent In any proposal for the physical implementation of quantum computing, the final qubit readout plays an integral role\cite{david}. The same holds true for less ambitious forms of controlled quantum-state manipulation. This prompts the study of detector quantum efficiency: what is the relation between the minimum measurement time $\tau_{\rm m}$ and the rate $\Gamma_{\rm d}$ at which coherence in the measured system is destroyed? It turns out (e.g.,\cite{eta-eq}) that the fundamental limit reads $\eta\equiv(2\tau_{\rm m}\Gamma_{\rm d})^{-1}\le1$; in general, $\eta$ is called the detector nonideality. While the detector's decohering backaction noise thus is unavoidable, it need not be unknowable: if it is correlated with the detector's output noise, part of it is contained in the measurement record, limiting information loss. Accounting for this leads to a modified definition $\tilde{\eta}$ (cf.\cite{eta-eq} and Eq.~(\ref{def-eta}) below), with $\eta\le\tilde{\eta}\le1$. A formulation (also used in quantum optics) in terms of the energy sensitivity $\epsilon$ (cf.\ Eq.~(\ref{def-eps}) below), quantum-limited to $\epsilon\ge\half$ (throughout $\hbar=1$), leads to a similar conclusion\cite{danilov,dima}.

In solid-state realizations, the natural discreteness of either single electrons or especially Cooper pairs makes charge qubits one of the prominent candidates, and indeed experimental progress is encouraging\cite{nakamura}. The above then leads one to study {\em quantum detection of charge}. A point-contact electrometer was shown in Ref.\cite{K&A} to have $\eta\uparrow1$ for suitable parameters. While this is significant theoretically, the required 2-d electron gas and the limited sensitivity in current experiments make implementation difficult. A single-electron transistor (SET) is more attractive, setting forth the theoretical challenge to analyze detector efficiency in the presence of the strong Coulomb interaction typical of these devices. A major distinction must first be made between superconducting\cite{zorin} and normal SETs. The former have the advantages of straightforward integration into a circuit containing Josephson qubits, and of the absence of decohering normal metal\cite{shunt}. On the other hand, the latter will be robust against magnetic fields if these are e.g.\ needed to manipulate spins\cite{kane}; furthermore, they are sometimes preferred empirically, perhaps because the detailed behaviour of voltage-biased superconducting SETs is rather complicated\cite{SSET}. Without arguing in favour of either device, these remarks motivate the theoretical analysis of normally conducting SET quantum electrometers.

Such an analysis was carried out in Ref.\cite{shnirman} for semiclassical `orthodox' sequential tunneling. Although the detailed formulas apparently cannot be taken literally (cf.\cite{eta-eq}), the main conclusion is that $\tilde{\eta}={\cal O}(\alpha)\ll1$ for bias voltages $V$ well beyond the Coulomb-blockade threshold $V_{\rm t}$; here, $\alpha=(2\pi e^2\hn R_{\rm T})^{-1}$ is the dimensionless tunneling conductance. Only if $V\downarrow V_{\rm t}$ does $\tilde{\eta}$ increase to become of the order of unity in the crossover region, where the analysis breaks down. While thus not leading directly to quantum-efficient detection, the result encourages one to look beyond the quasi-Ohmic regime. An analysis for $V\approx V_{\rm t}$ was presented in Ref.\cite{dima} for a `quantum-dot' SET, with a central island consisting of a single energy level---the extreme opposite of the more widely used metallic SETs. With $\epsilon>1/\sqrt{3}$, ideal detection is then found to be unattainable.

Clearly, it is desirable to complement these studies by calculating $\tilde{\eta}$ for a metallic SET in the threshold and, for still smaller $V$, cotunneling regimes. For an SET, $\tilde{\eta}$ can be elegantly expressed in terms of conventional transport quantities as\cite{eta-eq}
\beq
  \tilde{\eta}=\frac{(C^2\!/4)(dI_n^{\rm st}\!/dq)^2
                     +\tilde{\cal F}_{\!Qn}(0)^2}
                    {{\cal F}_{\!nn}(0)\hp{\cal F}_{\!QQ}(0)}\;.
\eeql{def-eta}
Here, $dI_n^{\rm st}\!/dq$ is the responsivity of the DC current to the offset determining the available island charge states as $Q=q+Ne$ ($N\in{\bf Z}$), and $C$ is the total island capacitance. The current power spectrum ${\cal F}_{\!nn'}(\omega)=\int\!dt\,e^{i\omega t}F_{nn'}(t)$ is obtained from the correlator $F_{nn'}(t)\equiv\langle I_n(t)I_{n'}\rangle$, with indices $n^{(\prime)}={\rm l,r}$ for the left and right leads respectively, and with the average $\langle\cdot\rangle$ taken with respect to the density matrix of the stationary current-carrying state. The other power spectra are similarly obtained from the charge correlator $F_{QQ}(t)\equiv\langle Q(t)\hp Q\rangle$ and cross-correlator $F_{Qn}(t)\equiv\langle Q(t)I_n\rangle$. In quantum mechanics the operator ordering matters, and one defines the symmetrized $\tilde{F}_{Qn}(t)\equiv\half\langle Q(t)I_n+I_nQ(t)\rangle=\re F_{Qn}(t)$, so that $\tilde{\cal F}_{\!Qn}(\omega)=\half[{\cal F}_{\!Qn}(\omega)+{\cal F}_{\!Qn}{(-\omega)}^*]$ is real for $\omega=0$. The latter is understood as the low-frequency limit, that is, the contribution $\propto\delta(\omega)$ of the average quantities is excluded. Equations~(\ref{F-tot})--(\ref{FQs}) below then show that (\ref{def-eta}) in fact is $n$-independent. Taking only its first term, i.e., ignoring the output--backaction cross-correlation, yields the corresponding expression for~$\eta$. In terms of the same correlators, the energy sensitivity is defined by
\beq
  \epsilon=\sqrt{\epsilon_I^\phs\epsilon_Q^\phs-\epsilon_{IQ}^2}\;,
\eeql{def-eps}
with $\epsilon_I={\cal F}_{\!nn}(0)/[C(dI_n^{\rm st}\!/dq)^2]$, $\epsilon_Q={\cal F}_{\!QQ}(0)/C$, and $\epsilon_{IQ}=\tilde{\cal F}_{\!Qn}(0)/[C\hp dI_n^{\rm st}\!/dq]$.

With the above purpose, we have studied SET fluctuations using a diagrammatic Keldysh technique\cite{Sch2} in the many-channel or `wide-junction' approximation, appropriate in the metallic case. Tunneling is described by a Hamiltonian coupling reservoirs at chemical potentials $\mL$, $\mR$, and $\mu_{\rm i}\equiv0$ in the left and right leads and on the island respectively, so that $eV=\mL-\mR$\cite{mLR}. The electrons interact only via their collective charge, and one introduces the Coulomb-energy differences
\beq
  \Delta_N=\frac{(Q{+}e)^2-Q^2}{2C}=\frac{e}{C}[q+(N{+}\half)e]\;.
\eeql{Delta}
For $q\approx-e/2$, one needs to take only the states $N=0,1$ into account, all others having a much higher energy. For instance, $\Delta_0/\abs{\Delta_{-1}}\approx0.1$ if $q=-0.4e$. Since $eV_{\rm t}\ll\abs{\Delta_{-1}}$, this picture remains quantitatively correct throughout the cotunneling regime\cite{mLR}. One is left with three relevant energies (besides the temperature $T$), but inspection of Ref.\cite{Sch2} shows that
\beq
  I_n^{\rm st}(\mL,\mR,\Delta_0)=I_n^{\rm st}(\mL{-}\Delta_0,\mR{-}\Delta_0)
\eeql{Irel}
so that in fact there are only two parameters, corresponding to the bias and gate voltages of the experiment. This property is general for the two-state model, and we will absorb $\Delta_0$ by a shift $\mu_n\mapsto\mu_n-\Delta_0$.

For $\omega=0$, the diagrammatic resummation carried out in\cite{Sch2} for $I_n^{\rm st}$ can also be performed for ${\cal F}_{\!nn'}$, leading to a unified treatment of the cotunneling, crossover, and quasi-Ohmic regimes for low damping $\alpha\lesssim0.1$ and finite $T\ll E_C\equiv e^2\!/2C$. Alternatively, one can obtain the complete ${\cal F}(\omega)$'s to leading order in $\alpha$ on either side of $V_{\rm t}$. While these results are of interest and will soon be reported\cite{long}, presently we show how a subset of them suffices to evaluate $\tilde{\eta}$, and in particular that ideal detection {\em can\/} be attained within the range of validity of our approach.

Let us focus on the cotunneling regime, characterized (besides by $T=0$ and $\alpha_n\ll1$ for the individual junction conductances) by the impossibility of a semiclassical cascade $\mL>0>\mR$ (or vice versa). Without loss of generality take $\mR<\mL<0$ so that $V>0$, while the $N=0$ charge state is the dominantly occupied one. The low-frequency regime is defined by $\abs{\omega}<\min(eV,\abs{\mL})$\cite{no-ll}. The diagrammatic rules of Ref.\cite{Sch2} then yield\cite{long,average}
\bea
  &&\frac{{\cal F}_{\rm ll}(\omega)}{2\pi e^2}=\aL^2\theta(\omega)
   \biggl\{\frac{\omega^2{-}2\mL^2}{\omega}\ln\biggl(1{-}\frac{\omega^2}{\mL^2}
    \biggr)-2\omega\biggr\}\nonumber\\*
   &&\quad{}+\aL\aR\biggl\{2(\mR{-}\mL{-}\omega)+(\omega{+}\mL{+}\mR)
    \ln\biggl(\frac{\mL{+}\omega}{\mR}\biggr)\biggr\}\;,\nonumber\\[2mm]
  &&\frac{{\cal F}_{\rm\!rr}(\omega)}{2\pi e^2}=\aR^2\theta(\omega)
   \biggl\{\frac{\omega^2{-}2\mR^2}{\omega}\ln\biggl(1{-}\frac{\omega^2}{\mR^2}
    \biggr)-2\omega\biggr\}\nonumber\\*
   &&\quad{}+\aL\aR\biggl\{2(\mR{-}\mL{-}\omega)+(\omega{-}\mL{-}\mR)
    \ln\biggl(\frac{\mR{-}\omega}{\mL}\biggr)\biggr\}\;,\nonumber\\[2mm]
  &&\frac{{\cal F}_{\!n\bar{n}}(\omega)}{2\pi e^2}=-\frac{\aL\aR}{\omega}
 \biggl\{\mL(\mR{-}\omega)\ln\biggl(\frac{\mR{-}\omega}{\mL}\biggr)\nonumber\\*
   &&\qquad{}+\mR(\mL{+}\omega)\ln\biggl(\frac{\mL{+}\omega}{\mR}\biggr)
     +\omega(\mR{-}\mL{-}\omega)\biggr\}
\eeal{F-tot}
($\bar{n}$: `the other lead' so that $\bar{\rm r}=\rm l$ etc). For $\omega\rightarrow0$ these reduce to the Schottky formula
\beq
  {\cal F}_{\!nn'}(0)=(-)^{\delta_{\bar{n}n^{\hn\prime}}}eI_{\rm r}^{\rm st}\;,
\eeql{stky}
\vspace{-10mm}
\beq
  I_{\rm r}^{\rm st}=
  2\pi e\aL\aR\{(\mL{+}\mR)\ln(\mL/\hn\mR)+2(\mR{-}\mL)\}>0\;.
\eeql{Ist}
However, this `pure shot noise' is non-white on the frequency scale set by the $\mu$'s\cite{1/f}. Equation~(\ref{Ist}) is a limiting case of the full three-state cotunneling theory\cite{A&O}: letting $E_2\rightarrow\infty$ for one of the latter's two excitation energies $E_{1,2}$, one finds agreement for $E_1=\abs{\mL}$. This is consistent with our assessment of the two-state approximation.

The relative minus sign of ${\cal F}_{\!n\bar{n}}(0)$ is a consequence of current conservation; this will be important in the following. Namely, the operator identity $\dot{Q}=-\sum_nI_n$ implies
\begin{mathletters}
\bea
  i\omega{\cal F}_{\!Qn}(\omega)&=&\sum_{n'}{\cal F}_{\!n'n}(\omega)\;,
    \label{FQr-def}\\
  \omega^2{\cal F}_{\!QQ}(\omega)&=&
    \sum_{nn'}{\cal F}_{\!nn'}(\omega)\in{\bf R}\;.
\eeal{FQQ-def}
\label{Fdef}\end{mathletters}%
Substituting the correlators of Eq.~(\ref{F-tot}), it is first of all gratifying that the $\omega\rightarrow0$ limits implied by Eq.~(\ref{Fdef}) actually do exist. One finds in detail
\begin{mathletters}
\bea
  \tilde{\cal F}_{\!Qn}(0)&=&0\;,\label{FQr-res}\\
  {\cal F}_{\!QQ}(0)&=&2\pi e^2\aL\aR\frac{(\mL{-}\mR)^3}{6\mL^2\mR^2}>0\;.
\eeal{FQQ-res}
\label{FQs}\end{mathletters}%
Inspection of Eq.~(\ref{F-tot}) shows that the vanishing of $\tilde{\cal F}_{\!Qn}(0)$ is caused by not only ${\cal F}_{\!nn}(\omega)$ but also ${\cal F}_{\!n\bar{n}}(\omega)$ being real (with no excitable levels available to the system in the frequency range and perturbation order considered), the latter not being dictated by symmetry (outside equilibrium, not all the standard relations hold).

Equation~(\ref{FQQ-res}) (and also the relation (\ref{FQQ-def}) between the full spectra) has been verified by a direct diagrammatic calculation of ${\cal F}_{\!QQ}$, with the advantage that there are no lead-index dependencies, while the $Q$ operator merely selects the excited charge state. However, since $Q={\cal O}(\alpha^0)$ while $I_n={\cal O}(\sqrt{\alpha})$, the advantage is offset by the leading cotunneling calculation being already fourth order (second order for ${\cal F}_{\!nn'}$). The additional energy denominators mean that the $\omega\rightarrow0$ limit still requires care, and are also the reason why Eq.~(\ref{FQQ-res}) has no logarithms. More physically, this causes the more dramatic divergence in the threshold limit $\mL\uparrow0$ than for the current (\ref{Ist}), so that lifetime corrections in the threshold regime proper should be especially interesting for the charge fluctuations\cite{average}.

Returning to Eq.~(\ref{def-eta}), Eqs.~(\ref{Delta}) (for $N=0$) and~(\ref{Irel}) show that $d_qI_{\rm r}^{\rm st}=-(e/\hn C)[\partial_{\mL}{+}\partial_{\mR}]I_{\rm r}^{\rm st}$, the rhs following from Eq.~(\ref{Ist}). The detector nonideality now becomes a universal function of $\chi\equiv\mL/\hn\mR$,
\bea
  \tilde{\eta}=\eta&=&\frac{3[1+2\chi\ln\chi-\chi^2]^2}
            {2(\chi-1)^3[(1{+}\chi)\ln\chi+2(1{-}\chi)]}\label{eta-res}\\[3mm]
  &\sim&\left\{\begin{array}{ll} 3/[2\ln\chi^{-1}-4]\downarrow0\;,
         & \chi\downarrow0\;; \\
    1-\frac{1}{20}(1{-}\chi)^2+{\cal O}[(1{-}\chi)^3]\;, & \chi\uparrow1\;.
  \end{array}\right.\nonumber
\eea
Equation~(\ref{eta-res}) is this paper's central result. Its striking asymptotics have two significant consequences. First, $\tilde{\eta}$ deviates from its equilibrium value of one only slowly, so that even at $\chi=\half$ the nonideality is less than 2.5\%. This is useful, since the detector must be operated at nonzero bias. Second, for any reasonable $\alpha$'s, $\tilde{\eta}$ still differs appreciably from zero at the point $\chi\approx\aR$ where threshold lifetime effects cause our approximation to break down. Pending further investigation, this is consistent with a smooth crossover to the low semiclassical values for $\tilde{\eta}$\cite{eta-eq,shnirman}. Since $\tilde{\cal F}_{\!Qn}(0)=0$, Eq.~(\ref{def-eps}) yields simply $\epsilon=1/(2\sqrt{\eta})$.

The finding of detector ideality at small $V$ can be interpreted as follows: far away from $V_{\rm t}$, the SET's internal structure becomes invisible to the outside, so that the device behaves similarly to a tunable-barrier model\cite{goan}, which is known to be ideal. Furthermore, at $T=0$ and $V\rightarrow0$ the phase-space volume available for cotunneling processes shrinks to zero; hence, each process probes the detected charge almost identically, minimizing information loss\cite{private}. In the corresponding small-bias regime of the model in Ref.\cite{dima}, the limitation to one transport channel makes the output current impracticably low.

Let us finally investigate the `output' energy sensitivity $\epsilon_I$ [cf.\ below Eq.~(\ref{def-eps})], sometimes considered instead of---or indeed confused with---the total energy sensitivity~$\epsilon$. In particular, $\epsilon_I$ (which does not involve the backaction fluctuations) is the appropriate figure of merit only when the detected signal, coming from an external classical source, can be considered as a given. Equations (\ref{stky}) and~(\ref{Ist}) yield
\beq
  \epsilon_I=\frac{eV_{\rm t}}{4\pi\aL\aR E_C}
             \frac{\chi^2[(\chi{+}1)\ln\chi^{-1}+2(\chi{-}1)]}
                  {[1+2\chi\ln\chi-\chi^2]^2}\;,
\eeql{epsI}
where we used $eV_{\rm t}=\abs{\mR}$. Equation~(\ref{epsI}) diverges for\linebreak $\chi\uparrow1$, so that in terms of $\epsilon_I$ the detection actually is poor for small voltages. In the opposite limit $\chi\approx\aR$ [cf.\ below Eq.~(\ref{eta-res})], one finds $\epsilon_I\sim(eV_{\rm t}/4\pi E_C)\*(\aR/\aL)\*[\ln(\aR^{-1})-2]$. Since one can achieve $eV_{\rm t}\ll E_C$ by decreasing $\abs{q/\hn e+\half}$ (with an interesting possibility for further optimization by taking $\aR<\aL$), $\epsilon_I$ has no fundamental lower limit. These findings are in broad agreement with the semi-quantitative treatment in Ref.\cite{KALV}. For our purpose, Eqs.\ (\ref{eta-res}) and~(\ref{epsI}) moreover show that with $\chi\approx\half$, one can simultaneously achieve an excellent $\epsilon$ and a reasonable~$\epsilon_I$.

It would be interesting to go beyond cotunneling theory by extending the resummation calculation of ${\cal F}_{\!nn'}(0)$\cite{long} to the other correlators figuring in Eq.~(\ref{def-eta}). Even though a treatment of the crossover regime is rendered less urgent by our finding that rather $\tilde{\eta}\uparrow1$ in the opposite limit, one can then study the presumably degrading effects of finite temperature and quasiparticle damping. However, the perturbative findings outlined below Eq.~(\ref{FQQ-res}) caution one that taking the low-frequency limit may be difficult---in fact, it is delicate already for the current correlator. Foregoing both finite-damping and finite-frequency effects, one should also try to do a full three-charge-state cotunneling calculation\cite{A&O}. Our limitation to $\abs{q/\hn e+\half}\lesssim0.1$ leaves a usable finite window for operation if $\alpha\ll1$, but is at present purely technical. In particular, the question whether ideal quantum detection can be achieved for small $V$ and {\em general\/} $q$ remains open.

I thank I.L. Aleiner, D.V. Averin, M.H. Devoret, B.E. Kane, A.N. Korotkov (who also commented on the manuscript), K.K. Likharev, J.E. Mooij, Y. Nakamura, A.~Shnirman, and A.B. Zorin for discussions. D.V. Averin is furthermore gratefully acknowledged for introducing me to this topic, and for partial support through USAFOSR Grant 431-3999C. My stay in Stony Brook is facilitated by the Netherlands Organization for Scientific Research (NWO).


\end{multicols} 
 

\begin{references}

\bibitem{david} D.V. DiVincenzo, quant-ph/0002077.

\bibitem{eta-eq} A.N. Korotkov, cond-mat/0008461.

\bibitem{danilov} V.V. Danilov, K.K. Likharev, and A.B. Zorin, IEEE Trans. Magn. {\bf19}, 572 (1983).

\bibitem{dima} D.V. Averin, quant-ph/0008114.

\bibitem{nakamura} Y. Nakamura, Yu.A. Pashkin, and J.S. Tsai, Nature {\bf398}, 786 (1999).

\bibitem{K&A} I.L. Aleiner, N.S. Wingreen, and Y. Meir, Phys. Rev. Lett. {\bf79}, 3740 (1997); A.N. Korotkov and D.V. Averin, cond-mat/0002203.

\bibitem{zorin} A.B. Zorin, Phys. Rev. Lett. {\bf76}, 4408 (1996).

\bibitem{shunt} However, the total device of Ref.\cite{zorin}, while having $\eta=\tilde{\eta}\uparrow1$, does involve substantial on-chip normal metal in the form of a low-resistance shunt.

\bibitem{kane} B.E. Kane, Nature {\bf393}, 133 (1998).

\bibitem{SSET} E.g., A. Maassen van den Brink, G. Sch\"on, and L.J. Geerligs, Phys. Rev. Lett. {\bf67}, 3030 (1991); A. Maassen van den Brink, A.A. Odintsov, P.A. Bobbert, and G.~Sch\"on, Z.~Phys. B {\bf85}, 459 (1991).

\bibitem{shnirman} A. Shnirman and G. Sch\"on, Phys. Rev. B {\bf57}, 15400 (1998); see also M.H. Devoret and R.J. Schoelkopf, Nature {\bf406}, 1039 (2000).

\bibitem{Sch2} H. Schoeller and G. Sch\"on, Phys. Rev. B {\bf50}, 18436 (1994).

\bibitem{mLR} Electrostatics\cite{SSET} yields $\mL=eV(C_{\rm r}{+}C_{\rm g}/2)/C$, $\mR=-eV(C_{\rm l}{+}C_{\rm g}/2)/C$, where $C_{\rm l,r}$ are the junction capacitances, and where in practice the gate capacitance $C_{\rm g}$ can be neglected or absorbed into the other ones; $C=C_{\rm l}+C_{\rm r}+C_{\rm g}$. The threshold condition $\mL=\Delta_0$ becomes $eV_{\rm t}=\Delta_0C/(C_{\rm r}{+}C_{\rm g}/2)$, so that the estimate below Eq.~(\ref{Delta}) is correct for not too asymmetric SETs.

\bibitem{long} A. Maassen van den Brink, in preparation.

\bibitem{no-ll} Note that the inequality is not $\ll$; it does {\em not\/} involve further approximations, and the other cases have been handled as well.

\bibitem{average} Since the contribution ${\cal F}_{nn'}(\omega)=\cdots+2\pi\*I_{n\vphantom{k}}^{\rm st}I_{\smash{n^{\hn\prime}}\vphantom{k}}^{\rm st}\*\delta(\omega)$ is ${\cal O}(\alpha^4)$ in the cotunneling regime, it is not found in the second-order calculation of Eq.~(\ref{F-tot}). In Ref.\cite{long} it does show up, and can be separated from the genuine low-frequency fluctuations in a controlled way. However, the direct calculation of ${\cal F}_{\!QQ}$ outlined below (\ref{FQQ-res}) gives a finite but incorrect result for the  term $\propto\delta(\omega)$. This is an understood and presently harmless artifact of naive perturbation theory, which implicitly assumes $\abs{\omega}\gg\sum_n\alpha_n\abs{\mu_n}$ by ignoring island relaxation\cite{long}. We have checked that the problem does {\em not\/} occur for ${\cal F}_{\!nn'}(0)$: the cotunneling limit of the full resummation calculation agrees with Eqs.~(\ref{stky}), (\ref{Ist}), where the limits are taken in the opposite order.

\bibitem{1/f} Thus, in practice there will be a frequency range where the intrinsic correlators (\ref{F-tot}), (\ref{Fdef}) already reach their $\omega=0$ limit, while $\omega$ is still high enough so that $1/\hn f$-noise is small.

\bibitem{A&O} D.V. Averin and A.A. Odintsov, Phys. Lett. A {\bf150}, 251 (1989).

\bibitem{goan} S.A. Gurvitz, Phys. Rev. B. {\bf56}, 15215 (1997); H.-S. Goan, G.J. Milburn, H.M. Wiseman, and H.B. Sun, cond-mat/0006333.

\bibitem{private} A.N. Korotkov, private communication.

\bibitem{KALV} A.N. Korotkov, D.V. Averin, K.K. Likharev, and S.A. Vasenko, in {\it Single-Electron Tunneling and Mesoscopic Devices}, edited by H. Koch and H. L\"ubbig (Springer-Verlag, Berlin, 1992), p. 45.

\end{references}
\end{document}